\documentclass{article}

\usepackage[preprint]{neurips_2026}

\usepackage[utf8]{inputenc}
\usepackage{natbib}
\usepackage[T1]{fontenc}
\usepackage{hyperref}
\usepackage{url}
\usepackage{booktabs}
\usepackage{amsfonts}
\usepackage{amsmath}
\usepackage{nicefrac}
\usepackage{xcolor}
\usepackage{graphicx}
\usepackage{multirow}
\usepackage{subcaption}
\usepackage{placeins}
\graphicspath{{figures/}}

\newif\ifshowchecklist\showchecklistfalse

\title{GORGO: Online Tuning for Cross-Region Network-Aware LLM Serving}

\author{
  Alessio Ricci Toniolo \\
  Modal Labs \\
  \texttt{atoniolo@andrew.cmu.edu} \\
  \And
  Rome Thorstenson \\
  ART \\
  \texttt{rome@arcadiaresearch.team} \\
  \And
  Abinaya Dinesh \\
  ART \\
  \texttt{abinaya@arcadiaresearch.team} \\
}

\begin{document}

\maketitle

\begin{abstract}
Increasingly, LLM inference services proxy client requests to engine replicas distributed globally.
Load-balancing policies must jointly account for
factors including KV-cache locality, replica load, and variable network latency when
optimizing for metrics like latency and TTFT.
However, existing systems only evaluate a subset of these factors in their cost
model, leading to uneven concentrations of load and KV-cache across replicas. We
present GORGO, a proxy architecture that holistically factors network latency,
prefill cost, and queueing delay using tunable parameters. Since open-source chat
datasets such as LMSYS-Chat-1M and WildChat-4.8M lack long-context, high
prefix-reuse data, we release a synthetic dataset, ART-Chat-2.5M, from
long-context production metadata. On a tuning window from ART-Chat-2.5M,
evolutionary strategies guide the GORGO policy's parameters to directly optimize
p95 TTFT. During held-out evaluation windows, we fix the parameter values learned
from tuning and improve p95 TTFT by 6.9--15.5\% and p95 end-to-end (E2E) latency by
14.3--30.9\% over baseline load-balancing policies such as simple session affinity
and prefix-cache.
The code and ART-Chat-2.5M dataset can be found at \href{https://github.com/atoniolo76/GORGO}{https://github.com/atoniolo76/GORGO}.
\end{abstract}

\section{Introduction}
\label{sec:intro}

In LLM serving systems, perceived latency to the user is dominated by the time-to-first-token (TTFT). On a single replica, TTFT is dominated by three costs: (i) prefill time, (ii) round trip time (RTT) from client (proxy) to replica, and (iii) queueing delay behind in-flight requests. Prefix-caching, which is enabled in inference engines SGLang \citep{sglang2024} and vLLM \citep{vllm2023}, eliminates the prefill cost of previous turns in a multi-turn conversation. As LLM context windows increase in length, the time saved by prefix caching 90\% of a prompt with 100,000 tokens reduces the prefill cost to 10,000 tokens and decreases TTFT substantially.

Since LLM deployments proxy requests to inference engines across regions, the cost savings of prefix-caching depend on choosing a replica with the request session's prefix. Popular routing policies such as consistent hashing and prefix reuse aim to distribute load evenly while creating affinity between a session's requests and replica(s) to maximize KV-cache reuse \citep{karger1997consistent, chord2003, sglang2024, skywalker}. In compute-constrained regimes, bursty workloads can saturate a high-affinity replica, causing head-of-line (HOL) queueing delays from decode memory contention and negating cost savings from prefix-cache reuse. Routing policies that holistically evaluate all costs related to TTFT can maintain high prefix-cache reuse while minimizing the negative effects of load saturation and heterogeneous network latency.

Existing load balancing policies such as least-load, session affinity, and prefix-reuse may account for replica load or KV-cache hit rate; however, no existing policies consider network latency in cross-region scenarios, which can range on the order of 10ms to 1s (Figure~\ref{fig:rtt}). GORGO's routing policy accounts for all three TTFT costs, normalizes the units of measurement via tunable parameters, and jointly optimizes parameters through online tuning on real user workloads. To tune and stress-test different routing policies, in (\S\ref{sec:characterization}) we compile an LLM traffic trace from real production requests with high prefix-reuse and long-context prompts. The trace follows Mooncake's FAST'25 format \citep{qin2024mooncake} containing per-request timestamps, which can be linearly scaled to simulate variable saturation profiles.

We benchmark GORGO on a series of user workloads from ART-Chat-2.5M, our sensitized production Mooncake trace, and sweep across variable time scales to effectively saturate replicas without simulating unrealistic HOL queueing delay. Over existing load balancing policy baselines, GORGO jointly balances optimal TTFT with request concentration across replicas. Under the continuous batching paradigm, the ES-driven hillclimb tuner exploits warmer replicas close to the proxy and dramatically reduces TTFT at the cost of end-to-end (E2E) latency. We characterize the trade-off between request concentration across replicas, which inflates E2E latency for unbalanced distributions, and TTFT. Our contributions help contextualize the performance of LLM proxy routing policies in real-world user workloads, and (\S\ref{sec:results}) lists a case-by-case scenario of when one would want to use aforementioned baseline policies, the online GORGO policy, and offline GORGO with held-out weights. Finally, we show how conditioning the GORGO cost model on proxy-recorded replica load mitigates subversion of TTFT via continuous batching and slashes request latency across a panoply of metrics.

\section{GORGO}
\label{sec:method}

\subsection{Design Space}

For LLM load balancing policies, the design space includes which routing signals to consider and methods for choosing environment settings.
 
\textbf{Routing Signals}: Across existing load balancing policies, two distinct routing signals are considered: (i) replica load and (ii) KV-cache locality.
Simple policies such as least-load and least-request aim to evenly balance replica load while prefix-cache and simple-session-affinity aim to maximize KV-cache reuse per request.
NVIDIA Dynamo~\citep{nvidiaDynamoRouter}, SGLang Router~\citep{sglang2024}, and Preble~\citep{preble} offer algorithms that jointly optimize for replica load and KV-cache locality.

GORGO takes inspiration from early content-delivery network design, which has many similarities to LLM serving.
\citet{wang2002cdn} used three signals---server load, network proximity, and cache locality---to reduce response time and increase capacity. Evaluating network proximity can improve
responsiveness, or TTFT, in geo-distributed serving environments. GORGO is the first LLM load balancing policy to also optimize (iii) network latency from client to server.

\textbf{Weight-selection Methods}: Methods for tuning the weights of a request scheduler fall into (i) offline and (ii) online protocols.
Existing offline methods include Dynamo's SLA Planner, which has a sophisticated method to autoscale prefill and decode workers to meet TTFT and ITL targets. 
It requires profiling before deployment to determine the optimal tensor parallelism configurations and scaling parameters for the planner to use during runtime~\citep{nvidiaDynamoPlanner}.

GORGO uses online calibration because cache state, queue state, and batching behavior are intrinsically influenced by the routing policy's decisions, and accurately replicating the conditions necessary to evaluate the policies requires a dedicated replay fleet, shadow execution, or a sufficiently accurate simulator of cache and continuous-batching dynamics.
After all, a request log generated under one policy does not contain the counterfactual cache and queue evolution that would have arisen under another set of weights.

\subsection{Cost Model}

TTFT is known to consist of three different costs: network latency, prefill time, and queueing delay \citep{banaserve2025}. The KV-Cache, which stores key-value pairs of input sequences, removes redundant prefill computation for user sequences sharing a prefix with previously computed sequences \citep{sglang2024}. In a cross-region deployment setting, for an inference engine replica $i$ holding a set of cached token prefixes $c_i$, the cost of TTFT can be defined as the following function, where $x_r$ is the input sequence of tokens for a request $r$ and the prefill time depends only on the set difference $(x_r \setminus c_i)$, the tokens in $x_r$ not already cached on $i$:

\begin{equation}
\mathrm{TTFT} = T_{\texttt{network}}(i) + T_{\texttt{queue}}(i) + T_{\texttt{prefill}}(x_r \setminus c_i)
\label{eq:ttft}
\end{equation}

Theoretically, $T_{\texttt{network}}$ and $T_{\texttt{queue}}$ correspond with round trip time from a client to server and the duration of request processing ahead of the incoming request $r$. However, inference engines support batching requests continuously in order to minimize waiting for completion of previous request processing \citep{orca2022}. While continuous batching allows admission of a new request into the currently running batch, the batch size is still bounded by a maximum number of concurrent requests and the available KV-cache budget, a critical knob that governs how much load a replica can admit before further requests wait in a queue \citep{vllm2023}.

In a distributed system, the temporal delay of retrieving queueing metrics from an engine replica makes cost evaluation challenging. We represent the input to $T_{\texttt{queue}}$ as the total number of tokens of requests without a completion event on the client proxy. $T_{\texttt{network}}$ is measured trivially as the exponential weighted moving average of a ping's round trip time from client proxy to server.
\begin{equation}
\mathrm{TTFT} = T_{\texttt{network}}(\mathrm{RTT}_i) + T_{\texttt{queue}}\left(i, \sum_{j=1}^{n_r} x_j\right) + T_{\texttt{prefill}}(x_r \setminus c_i)
\label{eq:ttft-proxy}
\end{equation}

\bigskip
\subsection{GORGO Proxy Design}

The GORGO proxy routes client requests to the engine replica with the minimum calculated cost from Equation~\ref{eq:ttft-proxy}, parameterized by weights $W_{\texttt{rtt}}$, $W_{\texttt{prefill}}$, and $W_{\texttt{queue}}$. These parameters weight the inputs $T_{\texttt{network}}$, $T_{\texttt{queue}}$, and $T_{\texttt{prefill}}$, which unfairly mixes the units of time and tokens, to normalize the correlated costs of latency, prefill cost, and queueing time on TTFT. The weight of $T_{\texttt{prefill}}$ is fixed to 1 because GORGO proxy makes routing decisions based on relative replica cost: only the ratio of weights matters in this design.
\begin{equation}
\mathrm{TTFT} = W_{\texttt{rtt}} * T_{\texttt{network}}(\mathrm{RTT}_i) + W_{\texttt{queue}} * T_{\texttt{queue}}\left(i, \sum_{j=1}^{n_r} x_j\right) + T_{\texttt{prefill}}(x_r \setminus c_i)
\label{eq:score}
\end{equation}

GORGO proxy uses a simple (1+1) evolutionary strategy to tune weights $W_{\texttt{rtt}}$ and $W_{\texttt{queue}}$ on the objective function, p95 TTFT.
Each parent weight $x_{t,k}$ is perturbed multiplicatively in log-space by a normal random variable $z_k$ times step size $\sigma$, and the new offspring weight $x_k'$ is clamped to values in the hyperparameter range $[lo_k, hi_k]$ where $lo_k > 0$ and $hi_k > 0$.

\begin{equation}
x_k' = \mathrm{clip}\left(\exp(\ln(x_{t,k}) + \sigma_t * z_k), [lo_k, hi_k]\right)
\end{equation}

When $x'$ beats the parent weight $x_t$ on the objective metric, the incumbent weight $x_{t+1}$ is updated to $x'$, and $\sigma$ is adjusted to maintain Rechenberg's 1/5 success rule~\citep{rechenberg1973} of roughly one accepted offspring for every five proposals.

\section{Dataset}
\label{sec:characterization}

\begin{figure}[t]
\centering
\includegraphics[width=\linewidth]{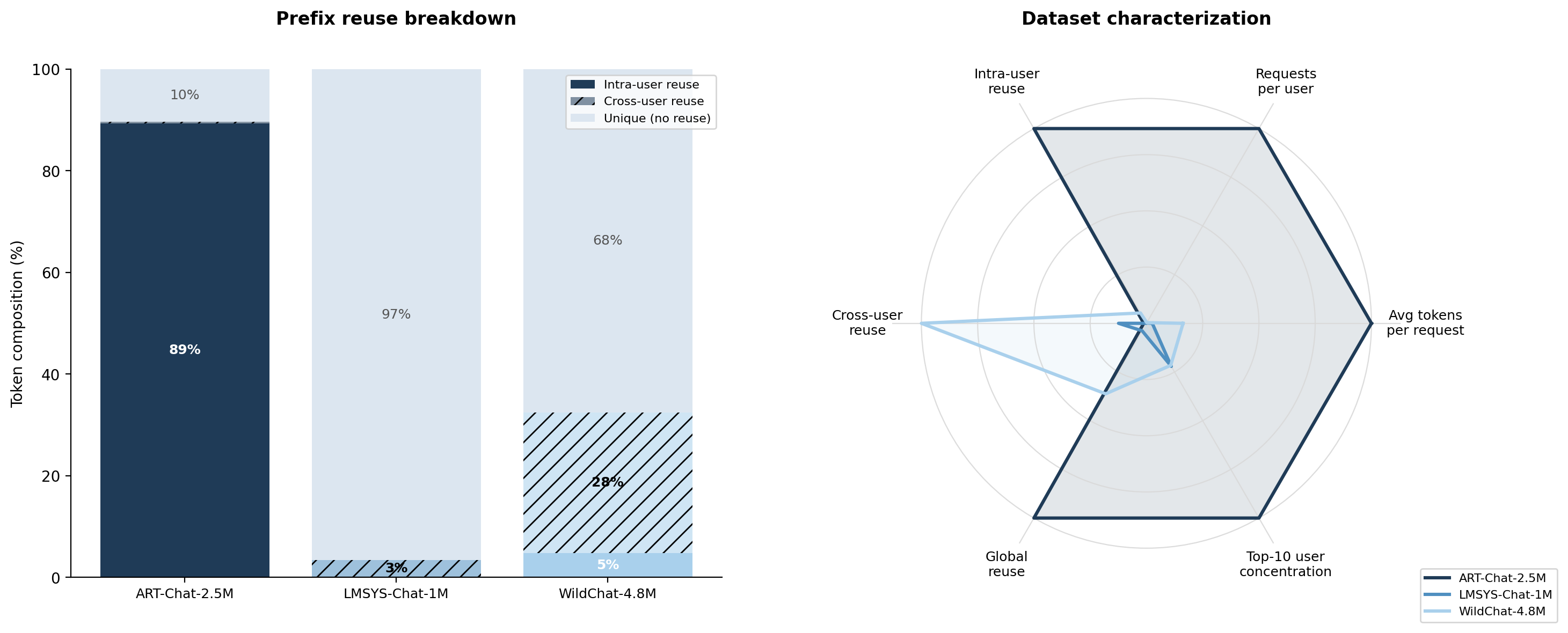}
\caption{\textit{Left}: Dataset characterization. ART-Chat-2.5M's prefix reuse
  is overwhelmingly intra-user (89\%) with minimal cross-user reuse (0.3\%), which reflects
  the long-context, multi-turn data; WildChat's prefix reuse is predominantly cross-user (28\%, shared templates); LMSYS
  has minimal cross-user reuse (3\%) and no intra-user reuse due to the lack of a field for client origin. \textit{Right}: ART-Chat-2.5M
  contains the maximum for each attribute except for cross-user reuse, which in WildChat is a function of the shorter average token length (2,925 tokens) and a common system prompt.}
\label{fig:dataset_combined}
\end{figure}

Existing LLM chatbot datasets lack two critical components for benchmarking
cache-aware policies: (i) prefill-bound requests with long-context prompts and
(ii) multi-turn workloads with high prefix-reuse between requests. For example,
we measure the average request length and global prefix reuse of
\textbf{LMSYS-Chat-1M}~\citep{lmsys2024} and \textbf{WildChat-4.8M}~\citep{wildchat2024},
two popular LLM datasets derived from public chatbot demos
(Table~\ref{tab:datasets}, Figure~\ref{fig:dataset_combined}). WildChat-4.8M
contains hashed IPs per request, allowing categorization of cross-user and
intra-user reuse while LMSYS-Chat-1M lacks user identification. Additional
results from benchmarking GORGO on WildChat-4.8M can be found in
Appendix~\ref{app:wildchat}.

\textbf{ART-Chat-2.5M} is a long-context, multi-turn dataset synthetically generated from
a week-long metadata trace of production inference traffic with the same
prefix-reuse structure as the original workload. We release a replay-ready trace
in the Mooncake FAST'25 format, which contains per-request timestamps, request
metadata, and synthetically generated chat completion data~\citep{qin2024mooncake}.
By storing request timestamps, one can linearly scale the time between requests
to control replica load. We characterize the dataset against WildChat-4.8M and
LMSYS-Chat-1M in Table~\ref{tab:datasets} and Figure~\ref{fig:dataset_combined}.
Notably, the intra-user prefix reuse and average token length in ART-Chat-2.5M
are 19$\times$ and 6$\times$ higher than in WildChat-4.8M.

\begin{table}[t]
\centering
\small
\caption{Dataset. ART-Chat-2.5M contains higher average input tokens and global prefix reuse, which is measured by
adding the intra-user reuse and cross-user reuse, over other chat datasets.
}
\label{tab:datasets}
\begin{tabular}{lrrrrr}
\toprule
Dataset & Total reqs & Users & Avg input tokens & Intra-user reuse & Global reuse \\
\midrule
LMSYS-Chat-1M & 1{,}000{,}000 & --- & 467 & --- & 3.4\% \\
WildChat-4.8M & 3{,}199{,}860 & 1{,}833{,}730 & 2{,}925 & 4.7\% & 32.5\% \\
ART-Chat-2.5M & 2{,}525{,}215 & 4{,}984 & 17{,}964 & 89.4\% & 89.7\% \\
\bottomrule
\end{tabular}
\end{table}

\FloatBarrier
\section{Experimental Setup}
\label{sec:setup}

\begin{figure}[t]
\centering
\includegraphics[width=\linewidth]{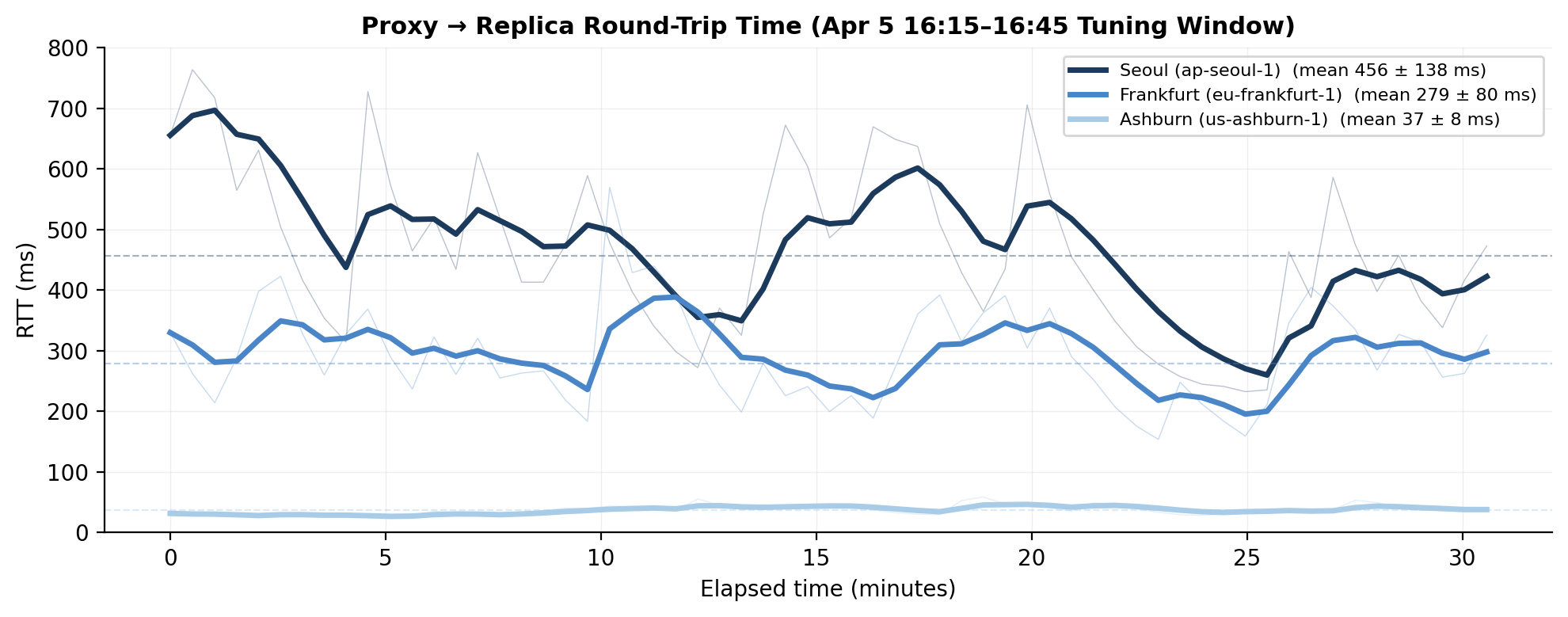}
\caption{Round trip time (RTT) between the GORGO policy's proxy in us-ashburn and the SGLang engines in us-ashburn, eu-frankfurt, and ap-seoul
over the Apr~5 16:15--16:45 tuning window. The proxy measures RTT every 30 seconds on a separate probe from the \texttt{/metrics} request and smoothes the RTT with an exponential moving average (EWMA).
Each region's mean latency demonstrates that network latency either adds a consideration when choosing between replicas with disparate load and prefill cost or simplifies the choice when choosing between replicas with similar costs.}
\label{fig:rtt}
\end{figure}

\paragraph{Proxy and engine configuration.}
Each policy runs the GORGO proxy on a small CPU worker in us-ashburn and controls
a dedicated SGLang inference engine in each of the following regions: us-ashburn,
eu-frankfurt, and ap-seoul. The engines contain two L40S GPUs each and serve the
Qwen3.5-35B-A3B model in FP8 format~\citep{qwen3technical}. All policies are
benchmarked on the same workload in parallel to rule out any variance in network
conditions. The round-trip time between the us-ashburn proxy and engines during
the tuning window is plotted in Figure~\ref{fig:rtt}. Due to the Qwen model's
limited context length of 32{,}768 tokens, we filter out any requests that
contain $>$24{,}000 tokens to leave adequate KV headroom. In SGLang, we set
\texttt{max\_concurrent\_requests} to 64 and \texttt{max\_output\_tokens} to 128
to limit unnecessary decode while simulating adequate load on the replica. All
workloads run alongside the proxy, dispatching requests to a local chat
completion endpoint.

\paragraph{Baseline policies.}
We benchmark the GORGO policy and compare performance of both online and static
modes to the below baselines. All SGLang metrics are scraped every 30 seconds
from the engine's Prometheus \texttt{/metrics} endpoint.
\begin{enumerate}
\item \texttt{least-load} minimizes the sum of proxy-tracked queued requests with
  SGLang metrics \texttt{num\_running\_reqs}, \texttt{num\_queue\_reqs}, and
  \texttt{num\_used\_tokens}. Queued requests to the proxy are defined as recently
  dispatched requests without a token response, and \texttt{num\_used\_tokens} are
  the currently occupied per-token KV slots.
\item \texttt{least-request} chooses the replica with the fewest in-flight
  requests from the proxy.
\item \texttt{prefix-cache} matches the request's prefix to the replica with the
  highest prefix-cache overlap, tracked on the proxy-side by a prefix trie of
  dispatched requests~\citep{aibrix2024}.
\item \texttt{simple-session-affinity} hashes the first 256 tokens from a request
  and routes to the replica with that prefix hash.
\end{enumerate}

\paragraph{Tuning and evaluation windows.}
We pick three 30-minute windows with high user diversity from the ART-Chat-2.5M
trace: Apr~5th 16:15--16:45, Apr~6 15:05--15:35, and Apr~7 19:45--20:15.
Statistics on each of these windows are found in Table~\ref{tab:windows}
(Appendix~\ref{app:windows}). We
assign Apr~5th as the window where we tune GORGO's weights online to minimize the
p95 TTFT of a rolling 128-request window with hop size 32. $w_{\mathrm{rtt}}$ and
$w_{\mathrm{queue}}$ are each initialized to 0.5 and 0.1 and restricted to ranges
$[0.05, 2.0]$ and $[0.05, 0.5]$. These values are hand-picked from the paradigm
of continuous batching in SGLang, where incoming requests can be scheduled into
the current batch, affecting TTFT less significantly than a fixed floor of
network latency between regions. The Apr~6--7 windows fix weight values GORGO
learned on the Apr~5 tuning window. Due to the greater number of requests in
Apr~6--7, we increase \texttt{time\_scale} to 2.0 and 3.0, respectively, to
control replica saturation (Table~\ref{tab:loadsweep}).

\FloatBarrier
\section{Results}
\label{sec:results}
\label{sec:results:p95}

\paragraph{Results across three windows.}

\begin{table}[htbp]
\centering
\small
\caption{Experiment results. In the tuning window, GORGO learns weights $w_{\mathrm{rtt}}{=}0.276$, $w_{\mathrm{queue}}{=}0.5$ after initializing
at weights $w_{\mathrm{rtt}}{=}0.5$, $w_{\mathrm{queue}}{=}0.1$. On held out evaluation windows, GORGO's weights are frozen, and
the policy outperforms every baseline policy on p95 TTFT.}
\label{tab:results}
\setlength{\tabcolsep}{2.7pt}
\begin{tabular}{lrrrrrrr}
\toprule
Policy & TTFT p50 (ms) & TTFT p95 & TTFT p99 & E2E p50 (s) & E2E p95 & E2E p99 & ITL (ms) \\
\midrule
\multicolumn{8}{l}{\textit{Tuning window --- Apr~5 16:15--16:45 (time\_scale${=}1.0$, 7{,}195 requests)}} \\
\midrule
\textbf{GORGO}          & \textbf{673} & 2{,}514          & 4{,}473          & \textbf{3.04}  & \textbf{8.91}  & \textbf{17.54} & \textbf{17.6} \\
simple-session-affinity & 712          & \textbf{2{,}428} & 5{,}190          & 3.48           & 18.27          & 22.61          & 20.2 \\
least-load              & 763          & 2{,}447          & \textbf{4{,}349} & 3.63           & 13.69          & 17.57          & 21.7 \\
least-request           & 968          & 3{,}970          & 7{,}012          & 4.37           & 16.62          & 20.81          & 25.2 \\
prefix-cache            & 1{,}477      & 6{,}784          & 9{,}613          & 13.14          & 22.63          & 26.81          & 82.3 \\
\midrule
\multicolumn{8}{l}{\textit{Held-out eval --- Apr~6 15:05--15:35 (time\_scale${=}2.0$, 7{,}630 requests)}} \\
\midrule
\textbf{GORGO}          & \textbf{491} & \textbf{1{,}584} & \textbf{2{,}222} & \textbf{1.80}  & \textbf{3.28}  & \textbf{4.37}  & \textbf{9.0} \\
simple-session-affinity & 627          & 1{,}875          & 3{,}056          & 2.01           & 4.75           & 7.34           & 10.3 \\
least-load              & 616          & 1{,}818          & 2{,}885          & 2.17           & 4.06           & 5.78           & 10.7 \\
least-request           & 634          & 1{,}852          & 2{,}484          & 2.08           & 3.99           & 5.23           & 9.9 \\
prefix-cache            & 574          & 1{,}798          & 2{,}750          & 2.07           & 4.95           & 10.34          & 10.6 \\
\midrule
\multicolumn{8}{l}{\textit{Held-out eval --- Apr~7 19:45--20:15 (time\_scale${=}3.0$, 8{,}663 requests)}} \\
\midrule
\textbf{GORGO}          & \textbf{398} & \textbf{1{,}417} & \textbf{2{,}061} & \textbf{1.74}  & \textbf{3.36}  & 6.81           & \textbf{9.3} \\
simple-session-affinity & 457          & 1{,}522          & 2{,}402          & \textbf{1.74}  & 3.92           & 7.12           & \textbf{9.3} \\
least-load              & 495          & 1{,}669          & 2{,}368          & 1.95           & 4.07           & \textbf{6.22}  & 10.0 \\
least-request           & 527          & 1{,}759          & 2{,}578          & 1.98           & 4.44           & 7.85           & 10.0 \\
prefix-cache            & 533          & 1{,}861          & 2{,}872          & 2.20           & 12.00          & 20.04          & 11.8 \\
\bottomrule
\end{tabular}
\end{table}

\begin{figure}[htbp]
\centering
\includegraphics[width=\linewidth]{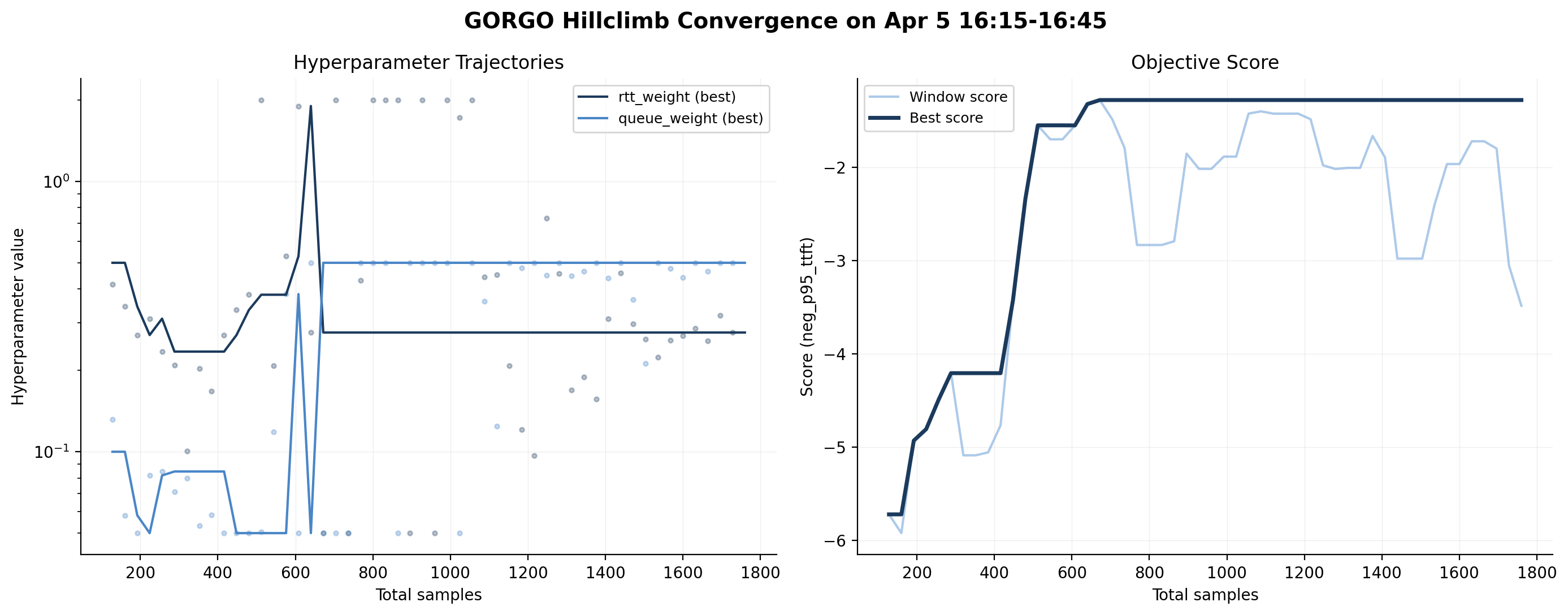}
\caption{Convergence of the evolutionary strategy on GORGO policy weights $w_{\mathrm{queue}}$ and $w_{\mathrm{rtt}}$.
Both weights reach their local optima after 672 samples, which occurs on the 18th evolutionary step. The fitness function is the negative p95 TTFT (in seconds) of the rolling 128-request tuning window, so a smaller magnitude is better; the best score reached is $-1.276$, i.e.\ a 1.276\,s best-window p95. Because it is computed over the rolling tuning window, this is lower than the 2{,}514\,ms overall tuning-window p95 in Table~\ref{tab:results}.}
\label{fig:es_convergence}
\end{figure}

Table~\ref{tab:results} reports TTFT, E2E latency, and inter-token latency (ITL)
for all policies in all three windows. While the GORGO policy's weights are
updated to minimize p95 TTFT during tuning, the held-out, fixed-weight evaluation
shows generalization of the learned values across days, with GORGO improving p95
TTFT by 6.9--15.5\% and E2E latency by 14.3--30.9\% over session-affinity. GORGO
slightly underperforms baseline policies on the tuning window because the
evolutionary strategy actively explores the space of parameters and tests worse weights than the learned solution, which converges after 672 samples
(Figure~\ref{fig:es_convergence}).

\paragraph{BFCL multi-turn replay.}
As an additional high-prefix-reuse workload, we replay 200 teacher-forced
episodes from BFCL's \texttt{multi\_turn\_base} subset, where each episode
contains sequential tool calls~\citep{patil2025bfcl}. We use two Qwen3-4B H200
replicas in us-east and us-west, 128 concurrent episodes, and at most 256 output
tokens per request. Identical prompts and tool results are replayed under GORGO
and session affinity.

\begin{table}[t]
\centering
\small
\caption{BFCL multi-turn replay. Bold marks the best value per column.}
\label{tab:bfcl}
\setlength{\tabcolsep}{2.7pt}
\begin{tabular}{lrrrrrrr}
\toprule
Policy & TTFT p50 (ms) & TTFT p95 & TTFT p99 & E2E p50 (s) & E2E p95 & E2E p99 & ITL (ms) \\
\midrule
\textbf{GORGO}          & \textbf{3{,}328} & \textbf{4{,}700} & 5{,}720          & \textbf{3.66} & \textbf{5.17} & \textbf{6.15} & --- \\
simple-session-affinity & 3{,}655          & 5{,}087          & \textbf{5{,}608} & 4.11          & 5.61          & 6.17          & --- \\
\bottomrule
\end{tabular}
\end{table}

Under high load, GORGO reduces p50 and p95 TTFT by 8.9\% and 7.6\%, and p50
and p95 E2E latency by 10.8\% and 7.9\%, respectively.

\paragraph{Load Sweep}

We find when sweeping across the \texttt{time\_scale} parameter that our SGLang inference engines reach a saturation point. After a concurrency threshold is reached,
requests begin to experience HOL queueing delay. In Table~\ref{tab:loadsweep}, the policies experience a 3$\times$ improvement in p95 TTFT and p95 E2E latency after \texttt{time\_scale=3.0} is reached.
In the tuning window, most policies other than GORGO are over saturating replicas due to the abnormally high p95 E2E latency when compared to the p95 E2E latency in later windows.
We recommend sweeping across timescales to find a clean under-saturated traffic profile before running GORGO. Due to the increased load profile of Apr 6-7, the timescale was increased to create a fair evaluation environment. However, we include in Appendix~\ref{app:apr7_ts2} a reference run of Apr~7 with \texttt{time\_scale}${=}2.0$ to show saturation at a lower timescale.

\begin{table}[htbp]
\centering
\small
\setlength{\tabcolsep}{4.5pt}
\caption{Load sweep. A sweep of the timescale variable, which linearly scales the time between requests, shows
p95 TTFT and E2E latency degrading at lower timescales. After \texttt{time\_scale${=}3.0$}, metrics improve by over 3$\times$ for unsaturated policies. Notably,
this timescale sweep shows that GORGO breaks the saturation point earlier than other policies due to a well-tuned load term.}
\label{tab:loadsweep}
\begin{tabular}{lrrrc}
\toprule
Policy & Input & TTFT p95 & E2E p95 & Saturated? \\
       & (tok/s) & (ms)     & (s)     &            \\
\midrule
\multicolumn{5}{l}{\textit{time\_scale${=}1.0$ (full load)}} \\
\midrule
gorgo                   & 34{,}064 & 8{,}260 & 15.76 & yes \\
simple-session-affinity & 34{,}035 & 1{,}835 & 17.94 & yes \\
least-request           & 34{,}030 & 6{,}138 & 18.62 & yes \\
\midrule
\multicolumn{5}{l}{\textit{time\_scale${=}2.0$}} \\
\midrule
gorgo                   & 17{,}003 & 1{,}822 & 5.65  & no  \\
simple-session-affinity & 16{,}999 & 1{,}707 & 15.50 & yes \\
least-request           & 16{,}999 & 4{,}361 & 16.85 & yes \\
\midrule
\multicolumn{5}{l}{\textit{time\_scale${=}3.0$}} \\
\midrule
gorgo                   & 11{,}327 & 1{,}378 & 3.25  & no  \\
simple-session-affinity & 11{,}326 & 1{,}556 & 4.04  & no  \\
least-request           & 11{,}326 & 1{,}805 & 4.92  & no  \\
\bottomrule
\end{tabular}
\end{table}

\paragraph{Exploiting Continuous Batching in SGLang}
\label{sec:results:hacking}
In the continuous batching paradigm, we learned that GORGO can stumble upon abnormal weight values and achieve an impressively low p95 TTFT at the cost of high p95 E2E latency.
Appendix~\ref{app:load_ablation} presents results from an experiment where the evolutionary strategy learns to set $w_{\mathrm{queue}}$
to 0 while keeping $w_{\mathrm{rtt}}$ near 0.23, which results in p95 TTFT 17\% better than the next-best policy. For this window,
GORGO chose to send 100\% of requests to the closest replica in us-ashburn. Continuous batching~\citep{orca2022} works by admitting incoming requests into the currently
running batch as long as some concurrency slot exists for the request. Thus, when all requests are sent to the same replica, any new requests sent to that replica can enter the batch, reuse existing KV-cache,
and prefill a single token. However, once the request enters the memory-bound decode phase, the replica's memory headroom saturates, KV-cache thrashes between GPU and CPU memory, and ITL/E2E latency collapses~\citep{orca2022, vllm2023}.


\section{Related Work}
\label{sec:related}

\paragraph{Prefix caches and reuse.}
RadixAttention in SGLang \citep{sglang2024} stores per-request KV state in a
radix tree; PagedAttention in vLLM \citep{vllm2023} enables prefix sharing
through paged memory. KVLink \citep{kvlink2025}, ChunkKV \citep{chunkkv2025},
KVFlow \citep{kvflow2025}, and Learned Prefix Caching \citep{lpc2025} improve
the single-replica cache but do not decide which replica serves a request.

\paragraph{Cross-replica routing and cross-region serving.}
Preble \citep{preble} introduces longest-prefix-match routing across replicas
with a load-balance fallback; AIBrix \citep{aibrix2024} packages a
production-grade variant of the same design and supplies the baseline we run.
Mooncake \citep{qin2024mooncake} is a KV-cache-centric architecture and the
source of our trace format. SkyServe \citep{skyserve} and SkyWalker
\citep{skywalker} address cross-region LLM serving at the placement and
spillover layers respectively, but treat per-request routing as out of scope.
k-LPM \citep{klpm2025} formulates LLM scheduling under TTFT constraints as
NP-hard. DLPM \citep{dlpm2025} targets fairness with locality. Llumnix
\citep{llumnix2024} migrates requests across replicas. CacheBlend
\citep{yao2024cacheblend} fuses cached KV state from reused non-prefix chunks
for retrieval-augmented generation, but does not address cross-replica routing
or network RTT. DistServe \citep{distserve2024} and Splitwise
\citep{splitwise2024} disaggregate prefill from decode. GORGO is the first
single-router policy in this space that scores cache locality, replica load,
and wide-area latency in one cost and derives the cost weights from the
deployment's own per-request TTFT stream, with no engine modifications.

\paragraph{Online hyperparameter adaptation.}
Unlike Bayesian-optimization or bandit approaches that maintain a surrogate
model, the $(1{+}1)$-ES \citep{rechenberg1973} provides fast, overhead-free
convergence in our 2-dimensional weight space.

\section{Limitations}
\label{sec:limitations}

Several constraints bound our conclusions. First, GORGO does not currently model heterogeneous hardware setups,
where replicas may have different prefill rates, queueing behavior, or KV-cache
capacities. Second, because the online
tuner optimizes p95 TTFT, it can exploit continuous batching by concentrating
load on the nearest replica, trading E2E and ITL tails for TTFT
(\S\ref{sec:results:hacking}). Adding a lower bound to the queue term mitigates but does not eliminate
this trade-off. Third, we do not consider prefill/decode disaggregation, which introduces
KV-transfer cost. Finally, GORGO does not manage or
directly observe replica-side KV-cache eviction. Consequently, the proxy's
cache index may temporarily overestimate prefix availability when a cached
prefix is evicted between routing and execution.

\section{Conclusion}
\label{sec:conclusion}

Routing across LLM replicas is a three-signal decision balancing cache locality, replica
load, and wide-area latency, but production heuristics commit to one signal and
degrade when that stops being the limiting factor. We treat the three as
terms of an additive per-replica cost and let online TTFT feedback optimize 
scaling weights, in place of operator-tuned constants or offline profiling. On a
long-context, high-prefix-reuse production trace the GORGO policy family
collectively achieves the lowest TTFT at every reported percentile across both
held-out evaluation windows. On a held-out WildChat window,
fixed-weight GORGO improves p95 TTFT by 17.2\% and p99 TTFT by 37.6\% over
random routing (Appendix~\ref{app:wildchat}). On paired BFCL multi-turn
replays, it reduces p95 TTFT and E2E latency by 7.6\% and 7.9\% relative to session
affinity. The design choices of GORGO transfer to
deployments where the workload regime is not known in advance and a dedicated
calibration window before each redeploy is not affordable, making it effective in production LLM serving on long context, distributed workloads. 

\bibliographystyle{plainnat}
\bibliography{references}

@inproceedings{sglang2024,
  title     = {{SGLang}: Efficient Execution of Structured Language Model Programs},
  author    = {Zheng, Lianmin and Yin, Liangsheng and Xie, Zhiqiang and Sun, Chuyue and Huang, Jeff and Yu, Cody Hao and Cao, Shiyi and Kozyrakis, Christos and Stoica, Ion and Gonzalez, Joseph E. and Barrett, Clark and Sheng, Ying},
  booktitle = {Advances in Neural Information Processing Systems (NeurIPS)},
  year      = {2024}
}

@inproceedings{vllm2023,
  title     = {Efficient Memory Management for Large Language Model Serving with {PagedAttention}},
  author    = {Kwon, Woosuk and Li, Zhuohan and Zhuang, Siyuan and Sheng, Ying and Zheng, Lianmin and Yu, Cody Hao and Gonzalez, Joseph E. and Zhang, Hao and Stoica, Ion},
  booktitle = {Proceedings of the 29th Symposium on Operating Systems Principles (SOSP)},
  year      = {2023}
}

@inproceedings{lmsys2024,
  title     = {{LMSYS-Chat-1M}: A Large-Scale Real-World {LLM} Conversation Dataset},
  author    = {Zheng, Lianmin and Chiang, Wei-Lin and Sheng, Ying and Li, Tianle and Zhuang, Siyuan and Wu, Zhanghao and Zhuang, Yonghao and Li, Zhuohan and Lin, Zi and Xing, Eric P. and Gonzalez, Joseph E. and Stoica, Ion and Zhang, Hao},
  booktitle = {International Conference on Learning Representations (ICLR)},
  year      = {2024}
}

@inproceedings{wildchat2024,
  title     = {{WildChat}: {1M} {ChatGPT} Interaction Logs in the Wild},
  author    = {Zhao, Wenting and Ren, Xiang and Hessel, Jack and Cardie, Claire and Choi, Yejin and Deng, Yuntian},
  booktitle = {International Conference on Learning Representations (ICLR)},
  year      = {2024}
}

@inproceedings{patil2025bfcl,
  title     = {The Berkeley Function Calling Leaderboard ({BFCL}): From Tool Use to Agentic Evaluation of Large Language Models},
  author    = {Patil, Shishir G. and Mao, Huanzhi and Yan, Fanjia and Ji, Charlie Cheng-Jie and Suresh, Vishnu and Stoica, Ion and Gonzalez, Joseph E.},
  booktitle = {Proceedings of the 42nd International Conference on Machine Learning (ICML)},
  pages     = {48371--48392},
  year      = {2025},
  volume    = {267},
  publisher = {PMLR}
}

@book{rechenberg1973,
  title     = {Evolutionsstrategie: Optimierung technischer Systeme nach Prinzipien der biologischen Evolution},
  author    = {Rechenberg, Ingo},
  year      = {1973},
  publisher = {Frommann-Holzboog},
  address   = {Stuttgart}
}

@misc{qwen3technical,
  title        = {{Qwen3} Technical Report},
  author       = {{Qwen Team}},
  year         = {2025},
  eprint       = {2505.09388},
  archivePrefix = {arXiv},
  primaryClass = {cs.CL}
}

@inproceedings{preble,
  title     = {{Preble}: Efficient Distributed Prompt Scheduling for {LLM} Serving},
  author    = {Srivatsa, Vikranth and He, Zijian and Abhyankar, Reyna and Li, Dongming and Zhang, Yiying},
  booktitle = {International Conference on Learning Representations (ICLR)},
  year      = {2025}
}

@misc{nvidiaDynamoRouter,
  title        = {{NVIDIA Dynamo}: Router Guide},
  author       = {{NVIDIA}},
  year         = {2026},
  howpublished = {\url{https://docs.nvidia.com/dynamo/latest/user-guides/kv-cache-aware-routing}},
  note         = {Accessed July 28, 2026}
}

@misc{nvidiaDynamoPlanner,
  title        = {{NVIDIA Dynamo}: Profiler},
  author       = {{NVIDIA}},
  year         = {2026},
  howpublished = {\url{https://docs.nvidia.com/dynamo/components/profiler}},
  note         = {Accessed July 28, 2026}
}

@inproceedings{wang2002cdn,
  title     = {The Effectiveness of Request Redirection on {CDN} Robustness},
  author    = {Wang, Limin and Pai, Vivek S. and Peterson, Larry L.},
  booktitle = {5th Symposium on Operating Systems Design and Implementation (OSDI)},
  year      = {2002},
  publisher = {USENIX Association},
  doi       = {10.1145/844128.844160}
}

@misc{aibrix2024,
  title        = {{AIBrix}: Towards Scalable, Cost-Effective Large Language Model Inference Infrastructure},
  author       = {{The AIBrix Team}},
  year         = {2025},
  eprint       = {2504.03648},
  archivePrefix = {arXiv},
  primaryClass = {cs.DC}
}

@inproceedings{distserve2024,
  title     = {{DistServe}: Disaggregating Prefill and Decoding for Goodput-optimized Large Language Model Serving},
  author    = {Zhong, Yinmin and Liu, Shengyu and Chen, Junda and Hu, Jianbo and Zhu, Yibo and Liu, Xuanzhe and Jin, Xin and Zhang, Hao},
  booktitle = {18th USENIX Symposium on Operating Systems Design and Implementation (OSDI)},
  year      = {2024}
}

@inproceedings{splitwise2024,
  title     = {{Splitwise}: Efficient Generative {LLM} Inference Using Phase Splitting},
  author    = {Patel, Pratyush and Choukse, Esha and Zhang, Chaojie and Shah, Aashaka and Goiri, {\'I}{\~n}igo and Maleki, Saeed and Bianchini, Ricardo},
  booktitle = {ACM/IEEE 51st Annual International Symposium on Computer Architecture (ISCA)},
  year      = {2024}
}

@misc{kvflow2025,
  title        = {{KVFlow}: Efficient Prefix Caching for Accelerating {LLM}-Based Multi-Agent Workflows},
  author       = {Wang, Zaifeng and Wei, Jiarui and Zhao, Pengfei},
  year         = {2025},
  eprint       = {2507.07400},
  archivePrefix = {arXiv},
  primaryClass = {cs.DC}
}

@inproceedings{lpc2025,
  title     = {Learned Prefix Caching for Efficient {LLM} Inference},
  author    = {Yang, Dongsheng and Li, Austin and Li, Kai and Lloyd, Wyatt},
  booktitle = {Advances in Neural Information Processing Systems (NeurIPS)},
  year      = {2025}
}

@misc{kvlink2025,
  title        = {{KVLink}: Accelerating Large Language Models via Efficient {KV} Cache Reuse},
  author       = {Yang, Jingbo and Hou, Bairu and Wei, Wei and Bao, Yujia and Chang, Shiyu},
  year         = {2025},
  eprint       = {2502.16002},
  archivePrefix = {arXiv},
  primaryClass = {cs.CL}
}

@misc{chunkkv2025,
  title        = {{ChunkKV}: Semantic-Preserving {KV} Cache Compression for Efficient Long-Context {LLM} Inference},
  author       = {Liu, Xiang and Tang, Zhenheng and Chen, Hong and Dong, Peijie and Li, Zeyu and Tang, Xiuze and Liu, Xuming and Liu, Xiaowen},
  year         = {2025},
  eprint       = {2502.00299},
  archivePrefix = {arXiv},
  primaryClass = {cs.CL}
}

@inproceedings{qin2024mooncake,
  title     = {Mooncake: A {KVCache}-centric Disaggregated Architecture for {LLM} Serving},
  author    = {Qin, Ruoyu and Li, Zheming and He, Weiran and Zhang, Mingxing and Wu, Yongwei and Zheng, Weimin and Xu, Xinran},
  booktitle = {23rd USENIX Conference on File and Storage Technologies (FAST)},
  year      = {2025},
  eprint       = {2407.00079},
  archivePrefix = {arXiv},
  primaryClass = {cs.DC}
}

@inproceedings{skyserve,
  title     = {{SkyServe}: Serving {AI} Models across Regions and Clouds with Spot Instances},
  author    = {Mao, Ziming and Xia, Tian and Wu, Zhanghao and Chiang, Wei-Lin and Griggs, Tyler and Bhardwaj, Romil and Yang, Zongheng and Shenker, Scott and Stoica, Ion},
  booktitle = {Proceedings of the Twentieth European Conference on Computer Systems (EuroSys)},
  year      = {2025}
}

@misc{skywalker,
  title        = {{SkyWalker}: A Locality-Aware Cross-Region Load Balancer for {LLM} Inference},
  author       = {Xia, Tian and Mao, Ziming and Kerney, Jamison and Jackson, Ethan J. and Li, Zhifei and Xing, Jiarong and Shenker, Scott and Stoica, Ion},
  year         = {2025},
  eprint       = {2505.24095},
  archivePrefix = {arXiv},
  primaryClass = {cs.DC}
}

@misc{klpm2025,
  title        = {{LLM} Query Scheduling with Prefix Reuse and Latency Constraints},
  author       = {Dexter, Gregory and Tang, Shao and Fatahi Baarzi, Ata and Song, Qingquan and Dharamsi, Tejas and Gupta, Aman},
  year         = {2025},
  eprint       = {2502.04677},
  archivePrefix = {arXiv},
  primaryClass = {cs.LG}
}

@misc{dlpm2025,
  title        = {Locality-aware Fair Scheduling in {LLM} Serving},
  author       = {Cao, Shiyi and Wang, Yichuan and Mao, Ziming and Hsu, Pin-Lun and Yin, Liangsheng and Xia, Tian and Li, Dacheng and Liu, Shu and Zhang, Yineng and Zhou, Yang and Sheng, Ying and Gonzalez, Joseph and Stoica, Ion},
  year         = {2025},
  eprint       = {2501.14312},
  archivePrefix = {arXiv},
  primaryClass = {cs.DC}
}

@inproceedings{llumnix2024,
  title     = {Llumnix: Dynamic Scheduling for Large Language Model Serving},
  author    = {Sun, Biao and Huang, Ziming and Zhao, Hanyu and Xiao, Wencong and Zhang, Xinyi and Li, Yong and Lin, Wei},
  booktitle = {18th USENIX Symposium on Operating Systems Design and Implementation (OSDI)},
  year      = {2024}
}

@inproceedings{yao2024cacheblend,
  title     = {{CacheBlend}: Fast Large Language Model Serving for {RAG} with Cached Knowledge Fusion},
  author    = {Yao, Jiayi and Li, Hanchen and Liu, Yuhan and Ray, Siddhant and Cheng, Yihua and Zhang, Qizheng and Du, Kuntai and Lu, Shan and Jiang, Junchen},
  booktitle = {Proceedings of the Twentieth European Conference on Computer Systems (EuroSys)},
  year      = {2025}
}

@inproceedings{orca2022,
  title     = {{Orca}: A Distributed Serving System for {Transformer}-Based Generative Models},
  author    = {Yu, Gyeong-In and Jeong, Joo Seong and Kim, Geon-Woo and Kim, Soojeong and Chun, Byung-Gon},
  booktitle = {16th USENIX Symposium on Operating Systems Design and Implementation (OSDI)},
  year      = {2022}
}

@misc{banaserve2025,
  title        = {{BanaServe}: Unified {KV} Cache and Dynamic Module Migration for Balancing Disaggregated {LLM} Serving in {AI} Infrastructure},
  author       = {He, Yiyuan and Xu, Minxian and Wu, Jingfeng and Hu, Jianmin and Ma, Chong and Shen, Min and Chen, Le and Xu, Chengzhong and Qu, Lin and Ye, Kejiang},
  year         = {2025},
  eprint       = {2510.13223},
  archivePrefix = {arXiv},
  primaryClass = {cs.DC}
}

@inproceedings{karger1997consistent,
  title     = {Consistent Hashing and Random Trees: Distributed Caching Protocols for Relieving Hot Spots on the World Wide Web},
  author    = {Karger, David and Lehman, Eric and Leighton, Tom and Panigrahy, Rina and Levine, Matthew and Lewin, Daniel},
  booktitle = {Proceedings of the Twenty-Ninth Annual ACM Symposium on Theory of Computing (STOC)},
  pages     = {654--663},
  year      = {1997}
}

@article{chord2003,
  title   = {{Chord}: A Scalable Peer-to-Peer Lookup Protocol for Internet Applications},
  author  = {Stoica, Ion and Morris, Robert and Liben-Nowell, David and Karger, David R. and Kaashoek, M. Frans and Dabek, Frank and Balakrishnan, Hari},
  journal = {IEEE/ACM Transactions on Networking},
  volume  = {11},
  number  = {1},
  pages   = {17--32},
  year    = {2003}
}

\newpage





\appendix

\section{Evaluation Window Characteristics}
\label{app:windows}

\begin{table}[h]
\centering
\small
\caption{Workload characteristics of the three tuning and held-out windows in
  Table~\ref{tab:results}.}
\label{tab:windows}
\begin{tabular}{lrrr}
\toprule
 & Apr5 tuning & Apr6 eval & Apr7 eval \\
 & 16:15--16:45 & 15:05--15:35 & 19:45--20:15 \\
 & time\_scale 1.0 & time\_scale 2.0 & time\_scale 3.0 \\
\midrule
Distinct users               & 579      & 606      & 540      \\
Requests / user              & 12.8     & 12.9     & 16.4     \\
Avg turns / conv.  & 30.3     & 27.7     & 21.5     \\
Multi-turn requests & 73.9\%  & 74.1\%   & 73.7\%   \\
Median input tokens          & 5{,}638  & 5{,}787  & 5{,}745  \\
Avg input tokens             & 6{,}877  & 7{,}008  & 7{,}148  \\
p95 input tokens             & 19{,}886 & 20{,}974 & 20{,}435 \\
Block global reuse           & 78.8\%   & 77.6\%   & 87.9\%   \\
Block intra-user reuse       & 78.5\%   & 77.1\%   & 87.4\%   \\
\bottomrule
\end{tabular}
\end{table}

\section{Apr~7 at \texttt{time\_scale}${=}2.0$}
\label{app:apr7_ts2}

\begin{table}[h]
\centering
\small
\setlength{\tabcolsep}{2.7pt}
\caption{Apr~7 19:45--20:15 held-out window at \texttt{time\_scale}${=}2.0$}
\label{tab:apr7_ts2}
\begin{tabular}{lrrrrrrr}
\toprule
Policy & TTFT p50 (ms) & TTFT p95 & TTFT p99 & E2E p50 (s) & E2E p95 & E2E p99 & ITL (ms) \\
\midrule
GORGO                   & 586          & 1{,}891          & 3{,}431          & 2.38          & \textbf{7.10}  & \textbf{12.58} & 13.4 \\
simple-session-affinity & \textbf{512} & \textbf{1{,}684} & \textbf{2{,}455} & \textbf{2.19} & 14.66          & 17.19          & \textbf{12.2} \\
least-load              & 633          & 2{,}158          & 3{,}753          & 2.69          & 10.39          & 16.19          & 15.5 \\
least-request           & 915          & 4{,}806          & 7{,}114          & 3.77          & 17.27          & 19.61          & 19.7 \\
prefix-cache            & 1{,}474      & 8{,}243          & 12{,}167         & 7.67          & 22.14          & 26.66          & 45.7 \\
\bottomrule
\end{tabular}
\end{table}

\section{WildChat replay}
\label{app:wildchat}

\begin{table}[htbp]
\centering
\small
\caption{Held-out WildChat-4.8M replay at \texttt{time\_scale}${=}2.0$
(5.7 requests/s open-loop), $c{=}64$, using one H100 replica in each of
us-west4, CANADA-2, and sines-2. Fixed-weight GORGO uses the weights learned on
the first window ($w_{\mathrm{rtt}}{=}2.0$ and
$w_{\mathrm{queue}}{=}0.104$). Each arm contains 20{,}422 completed requests
with zero failures. Bold marks the best value per column; ITL is the median.}
\label{tab:wildchat}
\setlength{\tabcolsep}{2.7pt}
\begin{tabular}{lrrrrrrr}
\toprule
Policy & TTFT p50 (ms) & TTFT p95 & TTFT p99 & E2E p50 (s) & E2E p95 & E2E p99 & ITL (ms) \\
\midrule
GORGO  & \textbf{215} & \textbf{506} & \textbf{728}   & 1.55          & \textbf{2.42} & \textbf{2.98} & 10.1 \\
random & 244          & 611          & 1{,}166         & \textbf{1.38} & 2.51          & 4.04          & \textbf{8.4} \\
\bottomrule
\end{tabular}
\end{table}

\section{Load Weight Adaptation}
\label{app:load_ablation}

\begin{table}[htbp]
\centering
\small
\caption{Reward hacking on the Apr~2 12:30--13:00 window. With $w_{\mathrm{queue}}{=}0$, the
tuned \texttt{gorgo-static} wins every TTFT percentile but is worst on the E2E and ITL tails,
because it routes $\sim$100\% of traffic to a single replica that saturates under load. Bold
marks the best value per column; ITL is the median.}
\label{tab:load_ablation}
\setlength{\tabcolsep}{2.7pt}
\begin{tabular}{lrrrrrrr}
\toprule
Policy & TTFT p50 (ms) & TTFT p95 & TTFT p99 & E2E p50 (s) & E2E p95 & E2E p99 & ITL (ms) \\
\midrule
\multicolumn{8}{l}{\textit{Held-out eval --- Apr~2 12:30--13:00 (time\_scale${=}1.0$, 3{,}071 requests)}} \\
\midrule
gorgo-static (hacked)            & \textbf{180} & \textbf{1{,}072} & \textbf{1{,}810} & 2.05          & 12.49         & 16.63         & 14.1 \\
simple-session-affinity          & 268          & 1{,}715          & 2{,}947          & 1.74          & 3.88          & 5.90          & 10.8 \\
least-load                       & 278          & 1{,}669          & 2{,}665          & 1.69          & 3.89          & 8.94          & 10.1 \\
least-request                    & 274          & 1{,}285          & 1{,}902          & \textbf{1.37} & \textbf{2.61} & \textbf{3.34} & 8.6 \\
random                           & 283          & 1{,}456          & 2{,}124          & 1.40          & 2.92          & 4.24          & \textbf{8.5} \\
prefix-cache                     & 288          & 1{,}468          & 2{,}374          & 1.56          & 3.59          & 7.12          & 9.4 \\
\bottomrule
\end{tabular}
\end{table}

\begin{figure}[th]
\centering
\includegraphics[width=\linewidth]{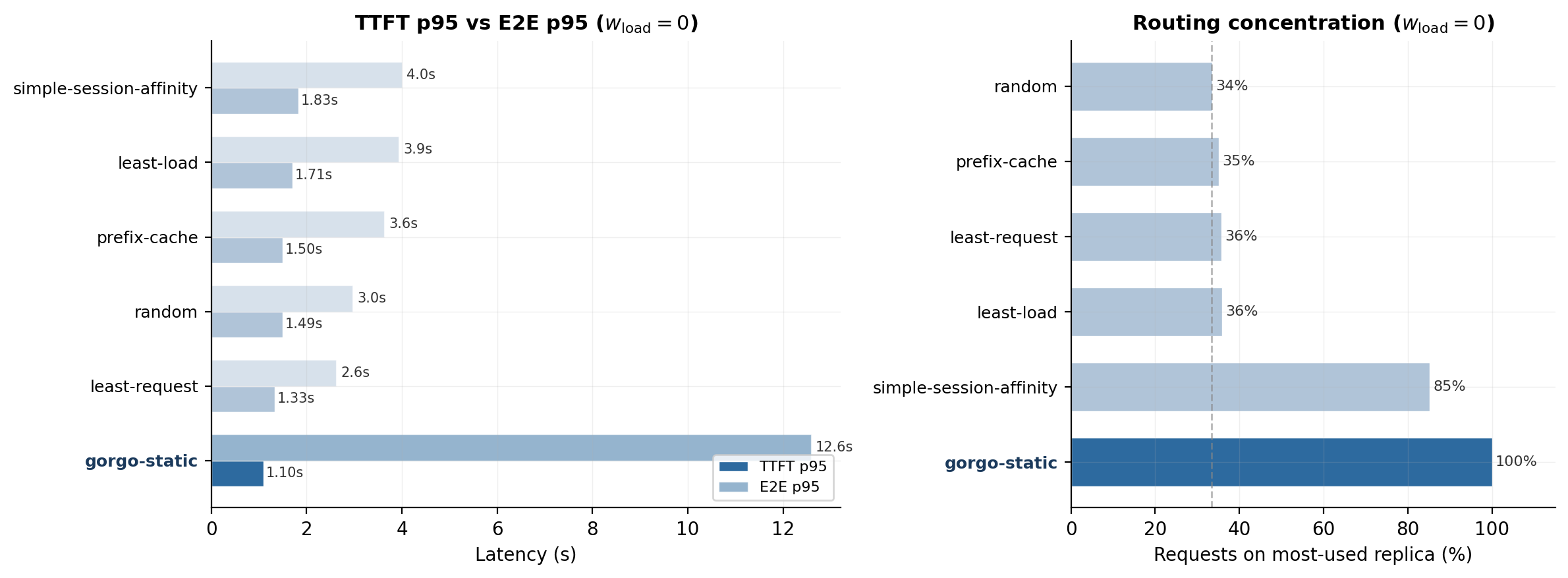}
\caption{\textit{Left}: TTFT p95 (dark) vs.\ E2E p95 (light) on the midday
  diurnal trace with $w_{\mathrm{queue}}{=}0$. \texttt{gorgo-static} wins
  TTFT but its E2E inflates to 12.6\,s. \textit{Right}: routing concentration.
  \texttt{gorgo-static} sends 100\% of requests to one replica.}
\label{fig:load_ablation}
\end{figure}

\end{document}